\documentclass[conference]{IEEEtran}
\usepackage{amssymb,amsmath}
\usepackage{amssymb,amsmath}
\usepackage{setspace}
\usepackage{graphicx}
\usepackage{epsfig,cite,latexsym,subfigure}
\usepackage{theorem}
\usepackage{times}
\usepackage{epsf,cite,subfigure}
\usepackage{amsmath,amssymb}

\usepackage{floatflt}
\usepackage{url}
\usepackage{times}
\usepackage[usenames]{color}
\usepackage{dsfont}
\usepackage{mathptmx}
\usepackage[final]{pdfpages}
\newcommand{\set}[1]{\left[#1\right]}
\newcommand     {\curlb}[1]{\left\{ #1\right\} }
\newcommand     {\paren}[1]{\left(#1\right)}
\newcommand{\C}{\mathbb{C}} 

\newcommand{\eqnlabel}[1]{\label{eqn:#1}}
\newcommand{\eqnref}[1]{(\ref{eqn:#1})}

\title{{\Large Reservoir Based Edge Training on RF Data\\ To Deliver Intelligent and Efficient IoT Spectrum Sensors}}\vspace{-2mm}
\author{
\IEEEauthorblockN{\normalsize Silvija Kokalj-Filipovic, Paul Toliver, William Johnson,  Rob Miller} \\ \vspace{-2mm}\\
\IEEEauthorblockA{\small Perspecta Labs Inc. \\
\small\em \{skfilipovic, ptoliver, wjohnson, rmiller\}@perspectalabs.com}}
\begin{document}

\maketitle
\begin{abstract}
Current radio frequency (RF) sensors at the 
Edge lack the computational resources to support
practical, in-situ training for intelligent spectrum monitoring,
and sensor data classification in general. We propose a solution
via Deep Delay Loop Reservoir Computing (DLR), a
processing architecture that supports general machine learning
algorithms on compact mobile devices by leveraging delay-loop
reservoir computing in combination with innovative electrooptical
hardware. With both digital and photonic realizations of our design of the loops,
DLR delivers reductions in form factor, hardware complexity and latency, compared to the State-of-the-Art (SoA). The main impact of the reservoir is to project the input
data into a higher dimensional space of {\em reservoir state vectors} in
order to linearly separate the input classes. Once the classes are
well separated, traditionally complex, power-hungry classification
models are no longer needed for the learning process. Yet,
even with simple classifiers based on Ridge regression (RR), the
complexity grows at least quadratically with the input size. Hence,
the hardware reduction required for training on compact devices
is in contradiction with the large dimension of state vectors.
DLR employs a RR-based classifier to exceed the SoA accuracy, while further
reducing power consumption by leveraging the architecture of
parallel (split) loops. We present DLR architectures composed of multiple smaller loops whose state vectors
are linearly combined to create a lower dimensional input into
Ridge regression. We demonstrate the advantages of using DLR
for two distinct applications: RF Specific Emitter Identification
(SEI) for IoT authentication, and wireless protocol recognition for IoT situational awareness.
\end{abstract}
\section{Introduction}
\begin{figure}[h]
\vspace{-1mm}
\centering
\hspace{-1mm}\includegraphics[width=0.5\textwidth]{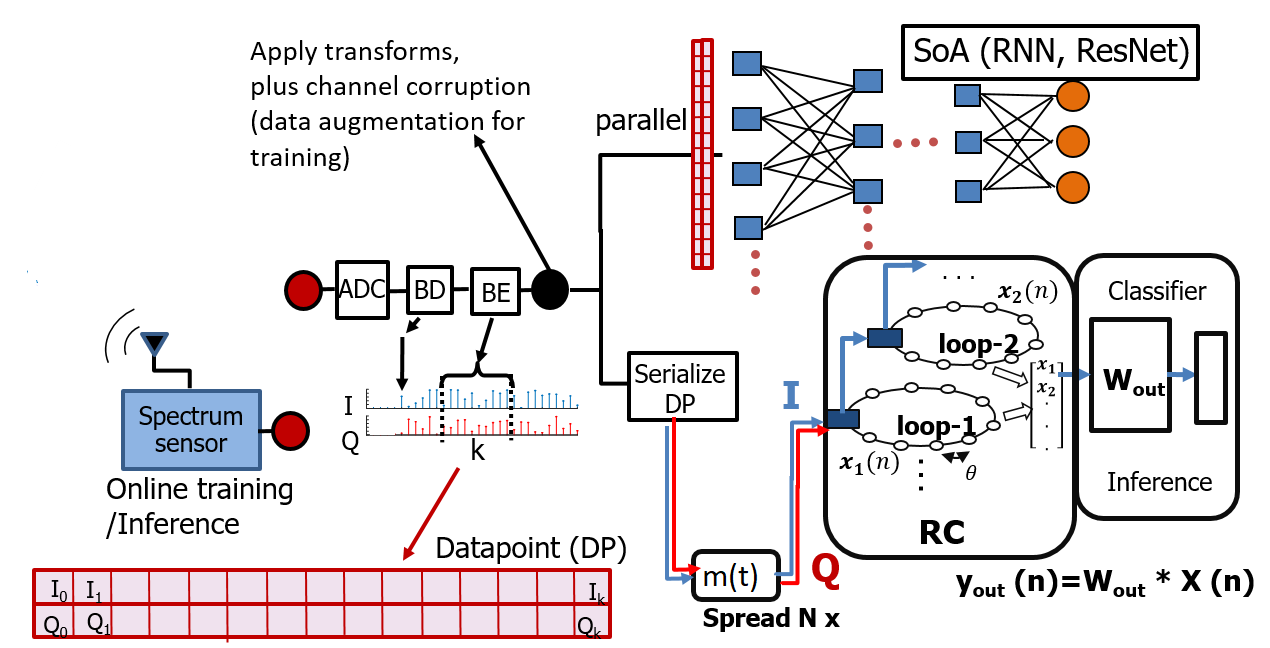}
\vspace{-3mm}
\caption{DLR system for SEI/WiPRec: BD is burst detection, while BE stands for burst extraction.
}\vspace{-3mm}
\label{fig:DLRsys}
\end{figure}

State-of-the-Art (SoA) machine learning systems that are trained on the sensor signals lack the computational resources to support in-situ training and adaptable inference for situational awareness. Such in-situ solutions are needed as it is not always practical to leverage backhaul resources due to security, bandwidth, and mission latency requirements. We propose a solution through Deep delay Loop Reservoir Computing (DLR), our novel AI processing architecture that supports general retrainable machine learning (ML) solutions on compact mobile devices by leveraging delay-loop reservoir computing (RC). Reservoir computing is a bio-inspired approach especially suited for processing time-dependent information in a computationally efficient way \cite{Lukosevicius2009ReservoirCA}. The RC in ML solutions conditions the input features towards linear separability of different classes, upon which any ML algorithm can be trained more efficiently.  DLR delivers significant reductions in form factor, hardware complexity/  power consumption for training at the Edge, providing real-time learning latency that is several orders of magnitude smaller than the SoA baseline classifiers. We demonstrate the advantages of DLR on the applications of RF Specific Emitter Identification (SEI) and wireless protocol recognition  (WiPRec). SEI  aims to extract rich nonlinear characteristics of internal components within a transmitter to distinguish one transmitter from another, even within the same manufacturer and protocol class. Complicated conventional authentication methods, based on the upper protocol layers and cryptography, are not convenient for small, computationally limited IoT devices. The authentication via RF-based SEI eliminates the communication and computation overhead in conventional approaches, while our approach makes it in-situ retrainable. For a pictorial illustration of the process employing a DLR system to train and perform SEI, including RF spectrum sensing and preprocessing, please consult  Fig.~\ref{fig:DLRsys}. 

We reported intermediate results in \cite{GomacTech}. Using architectural and algorithmic extensions,  we have since improved the DLR performance as well as the models of the SoA classifiers performing the SEI.  While we still utilize 2 deep neural net models for comparison, recurrent (RNN) and residual (ResNet), we managed to reduce their complexity  for a  fair comparison. It is difficult  to establish an SoA model for the collected dataset, which is not open source, but we made comparisons with SEI tasks of similar complexity. For example, the number of trainable parameters of a ResNet cited in \cite{hanna2020open} for the SEI on the set of 10 transmitters (plus an outlier class) is on the same order of magnitude as the ResNet we train as SoA reference on the dataset comprised of the signal bursts emitted from 20 WiFi devices.

Apart from the SEI dataset, we also evaluated DLR on the protocol recognition dataset comprised of the bursts of RF samples collected from four ISM emitter
classes, WiFi 802.11n \cite{80211n}, Bluetooth (BT) \cite{BT}, ZigBee \cite{ZBee}
and NRF \cite{NRF}, which we used in our prior publication \cite{SDAE}.
Our innovative architecture of split loops helps preserve the H/W reduction while exceeding the SoA accuracy in both applications. 

\begin{figure}[h]
\vspace{-1mm}
\centering
\includegraphics[width=0.49\textwidth]{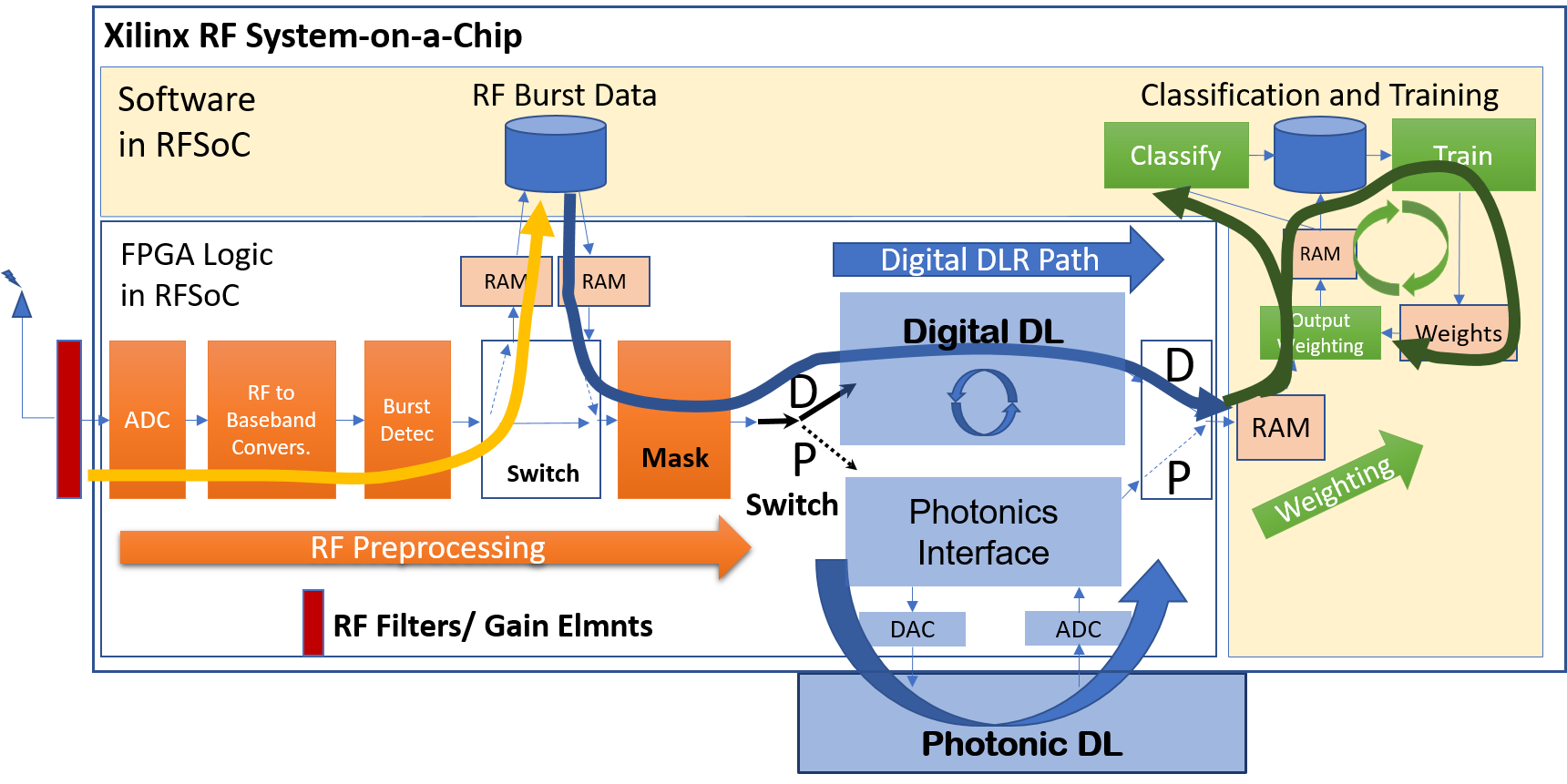}
\vspace{-3mm}
\caption{The DLR demo platform performing the SEI in-situ training and inference with two selectable delay loop implementations (D, for digital, and P for photonics. The yellow path is application specific (RF data ingest and burst extraction), while the blue path is generic across applications: burst of samples $~\longmapsto ~$ delay loops $~\longmapsto ~$ state vector $~\longmapsto ~$ RR training or classification}\vspace{-3mm}
\label{fig:platform}
\end{figure}

Our demo DLR platform \cite{RFSoC} supports two HW implementations of delay loops, using a configurable switch (see Fig.~\ref{fig:platform}): a digital (FPGA) loop, and an analogue one based on the innovative photonic hardware that exploits the inherent speed and multi-dimensional (spatial, temporal and wavelength-based) processing diversity of signals in the optical domain \cite{AdvancesinphotonicRC}. DLR may be configured to synergistically combine different input transformations with split loops, which offers flexibility for different ML applications and maintains the same outstanding performance. In its photonic implementation, this can be further combined with Wavelength Division Multiplexing (WDM) to process multiple split loops in parallel while preserving salient information. 

By showcasing 2 different datasets, we demonstrate that DLR supports a range of applications by adjusting its parameters and architectural combinations to achieve low SWaP, high accuracy and required latency. 
The organization of this paper is as follows: Section II describes motivation for the featured applications, and the existing work; Our algorithmic and architectural solutions based on the delay-loop concept are described in Section III; Section IV presents experimental results, including accuracy for different configurations  and H/W reduction figures for the SEI platform.

\section{Motivating Applications for DLR}
5G and open radio access networks (Open RANs) will result in hardware deployments that require additional efforts towards mitigating wireless security risks. As all electronic devices have fingerprints due to manufacturing variability, so do radio frequency emitters. SEI based on RF fingerprinting can individually identify a multitude of wireless devices \cite{CabricSEIAuthor}. SEI can be particularly useful in IoT, with billions of small devices with diverse cyber-security vulnerabilities including authentication and tracking \cite{PLASurvey}. The SEI based authentication is passive and secure: the fingerprint cannot be emulated to launch an attack like MAC address spoofing. However, the SoA approaches to SEI require big servers for training, and lack the in-situ training solutions. Previous attempts at in-situ training require extensive HW-specific pre-processing and have been evaluated only on simulated data \cite{PUF}. Hence the need for solutions like DLR.

Additionally, neural network (NN) based wireless receivers will soon become a reality. They implement various signal processing algorithms used in traditional receivers compliant to specific standards by using trained neural networks (see \cite{deepunfolding,modelbased} and references therein). To make them adaptable, we could use an intelligent spectrum sensor, such as DLR,  to identify the protocol of the incoming signal and feed it into a specifically trained stack of NNs. Our wireless protocol recognition (WiPReq) DLR would allow us to realize an AI-defined receiver that can communicate across all protocols utilized in the ISM band. 

\textbf{Existing DL Work} As for the prior work that applies delay loops to the above applications, our work is the first one to the best of our knowledge. There is some recent work where reservoir loops are used in the general area of wireless communications, e.g., for channel equalization \cite{EqualAntonic,UnifiedRC}. For the recent advances in physical reservoir computing, including applications, please see \cite{Tanaka2019RecentAI,Chembo}.

\section{Algorithmic Explorations} 
\begin{figure}[h]
\vspace{-1mm}
\centering
\includegraphics[width=0.49\textwidth]{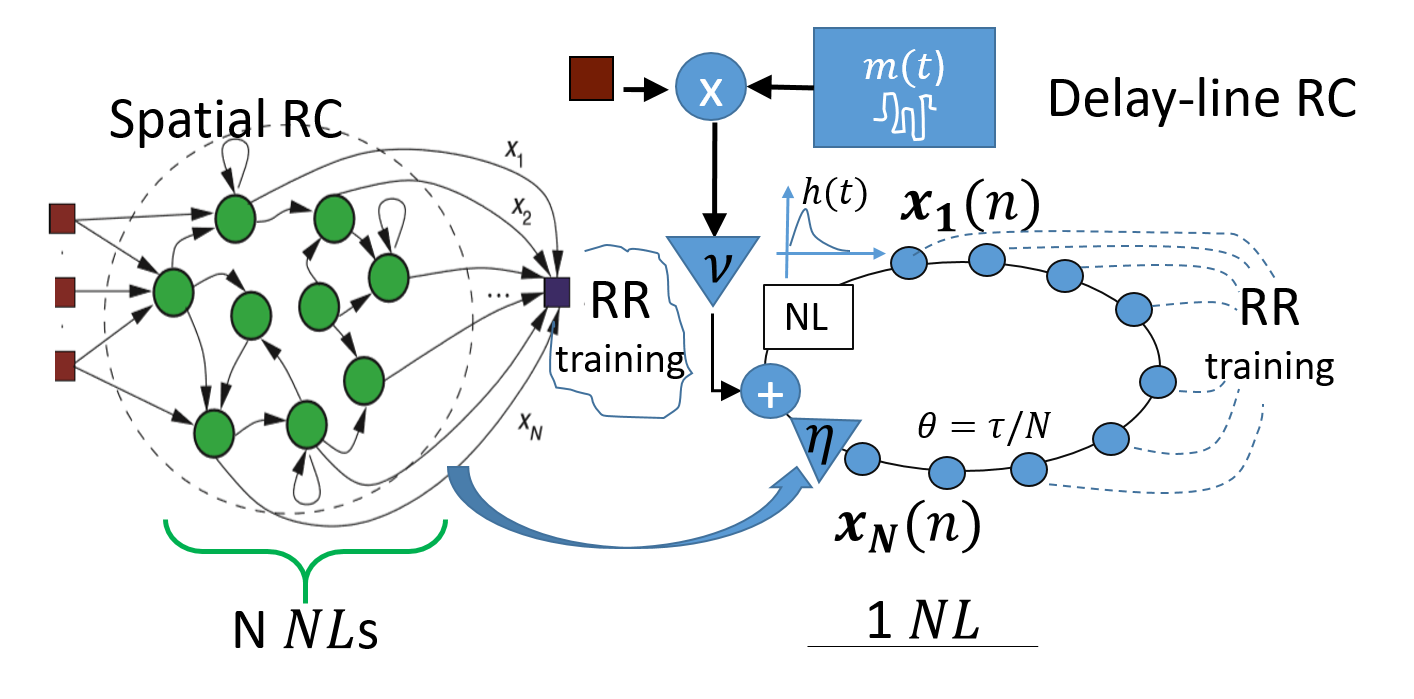}
\vspace{-5mm}
\caption{Delay-loop reservoir replaces the N neurons (NLs) on the left with a single one by using upsampling ($\theta$ depends on the loop bandwidth) and a simple design shown on the right}\vspace{-1mm}
\label{fig:basicdl}
\end{figure}

DLR uses reservoir computing in the form of a delay loop reservoir that replaces the N neurons in the traditional spatial implementation of the reservoir with N passes through a single neuron. The N-fold increase in delay by the sequential passing of the data through a single neuron is canceled out by an N-fold upsampling (sample and hold) done by a random spreading sequence $m(t)$ (see Fig.~\ref{fig:basicdl}). 
With reservoirs, the N must be large and such high rate upsampling is amenable to a photonic loop implementation. The wide bandwidth and wavelength diversity of photonic loops is the reason why DLR supports photonic DLs as a vehicle to future scalability. The N which allows the reservoir to linearly separate input classes for easier learning may be very large for certain applications. However, there are also other means of reducing the required reservoir size, which we present in this work.
In this paper we report results based on the FPGA and S/W (digital) implementations of the loop, as the emphasis is on algorithmic and architectural solutions. We are also emphasizing the training versus inference, as it is more demanding. Once trained the DLR maps the signal to its unique transmitter by just passing the burst of samples through the loop and multiplying it with the weight matrix. For a burst of 1 us, it happens in sub-millisecond including the extraction of the burst and its preprocessing by transforms. Figures of merit (FOMs) that we report for training include accuracy on the testing set, $H/W$ complexity, memory and latency.

\subsection{Dataset and Preprocessing}
The dataset to train the SEI detector now contains 20 classes, corresponding to 20 distinct WiFi devices. We first started with the dataset of 4 devices but it was not  sufficiently complex to allow for the exploration of the techniques that scale the design. To build this dataset we captured the emissions of commercial WiFi devices as they were sending beacons to an access point while using the same spoofed MAC address. 
We used USRP X310 with UBX RF daughterboard, with sample rate of 100 MHz centered in the middle of the 2.4 GHz ISM band. Our dataset is carefully prepared to not contain personally identifiable information. Datapoints are created by extracting bursts of 1024 complex (I/Q) samples from the captured time-series, right after the detection of the rising edge of the signal. Further preprocessing steps are explained in the subsection on transforms. 
\subsection{Delay Loop (DL)}

\begin{figure}[h]
\vspace{-1mm}
\centering
\includegraphics[width=0.39\textwidth]{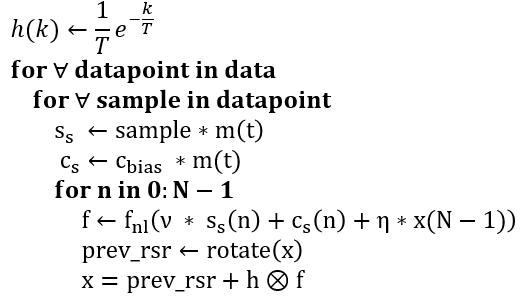}
\vspace{-4mm}
\caption{Pseudo code for the delay loop}\vspace{-1mm}
\label{fig:pseudo}
\end{figure}

The basic algorithm for the delay-loop state is expressed by~\eqnref{delayloop}.  
\begin{align}
&X_k (n)=\int_{\delta_k-\tau-\epsilon}^{\delta_k}{h_{\delta_k-\tau-\delta} f_{NL} \set{\eta X_\delta (n-1)+\nu J(n)(\delta + \tau)}d\delta.}  \eqnlabel{delayloop}
\end{align}
$X_k$ is the $k^{th}$ virtual element of the state vector $X.$ The upsampled time is defined in chips $\theta$ (in Fig.~\ref{fig:basicdl}), which gives rise to $J(n)(t),$ also denoted as $s_s(n)$ in the pseudo-code, Fig.~\ref{fig:pseudo}.  Each sample $s(n), n\in{1,\cdots,\ell}$ of the input datapoint of size $\ell$ (Fig.~\ref{fig:DLRsys}) is spread by the mask $m(t)$ and clocked into the loop as $J(n)(t),$ chip-by-chip. Here, $t \in {1,\cdots,N}$ is the chip-time index, and $k \in {1,\cdots,N}$ is the loop position index. Note that $N$ is the number of virtual reservoir nodes, as well as the length (in chips) of the spreading sequence (mask) $m(t)$ as shown in Fig.~\ref{fig:basicdl}.  $J(n),$ also indicated in Fig.~\ref{fig:platform}, is the input to the loop, and the output is read out after the last of the $\ell$ samples is clocked-in and put through the loop's non-linearity $N$ times.  

Each chip of the spread sample $J(n)(t)$ is linearly combined with the tail of $X: X_N(n-1)$ and put through the nonlinearity $NL.$  $X_N(n-1)$ has been affected by the same non-linearity at time $t-\tau,$ i.e., by the previous input sample $n-1$, where $\tau = N\theta$. Summation of the spread data input and  the tail of $X$ at every $t$ is practically creating the edges of the recurrent layer from the spatial implementation of the reservoir (the left side of Fig.~\ref{fig:basicdl})  \cite{AdvancesinphotonicRC}.  The randomness of edge weights is determined by the randomness of the spread sequence, which unfolds the edges at chip time. The output of the $NL$ may be convolved with filter $h(t)$ (Fig.~\ref{fig:pseudo}) to model the $NL's$ temporal response in photonics \cite{PhotRCTut},  which occurs on a finer time scale $d\delta$.  $\delta_k$ in the integral is the current time at node $k.$ 

We refer to the $X_k$ as the virtual node. No extra function or transformation is carried out on $X_k$ after the filtered non-linearity $f_{NL}.$ Omitting the propagation details, $X_k$s are simply time-shifted, in chip time $\theta,$ which also roughly matches the propagation time in photonics. \eqnref{delayloopD} models the shifting of the $f_{NL}$ output in digital implementaion, where $X_k$ at chip time $t$ is given by 
\begin{align}
\nonumber &X_k (t)=\\
 &\sum_{u=0}^{1}{h(u) f_{NL} \set{\eta X_k(t-N +u)+\nu J(t-k-u)}}  + \sigma,\  \sigma \longrightarrow 0. 
\eqnlabel{delayloopD}
\end{align}
The loop gain parameter $\eta$  and input gain $\nu$ must be calibrated to provide a proper dynamic state of the reservoir. $h(t)$ in \eqnref{delayloopD} is sampled at chip time.  
The results are based on $f_{NL}=sin\paren{\cdot},$  although we also used $tanh\paren{\cdot}$ as the neuron, with similar effects. The pseudo-code in Fig.~\ref{fig:pseudo} is mappable to the photonic implementation, which was important for the development and calibration of the photonic loop allowing the comparison with the baselines achieved with the digital-only loop \cite{PhotRCTut}. 

Another innovation is the introduction of “deep”, multi-layer architectures with both stacked and parallel (split) reservoir loops.  The rationale for the former is that splitting the input into $k$ parallel loops and combining their outputs reduces the complexity of the classifier by $k^2$ (please see section~\ref{subsec:split}). For the latter, we explored if $s$ sequential loops increase the representation complexity of the input allowing for more accurate classification. As the accuracy achieved to date by using sequential loops has not exceeded the one obtained with split loops only, and for space considerations, we here discuss just the split loops. We use transforms to adapt the salient signal information to the split loop input. We develop multiple ways to merge the outputs of the constituent DLR loops into a single output to be used to train a simple algorithm (Ridge Regression \cite{RR}). 
We used Bayesian hyperparameter optimization to find the best parameters, including the splitting. An idea is to treat the RC as a “black-box”, computationally heavy function: $f(\nu,\eta,N,s,k,\cdots)\rightarrow\set{0, 1},$ where the output is the resulting predictive accuracy.
Succinctly, the process involves the application of a Gaussian process prior on $f()$, then as more sets of parameters are sampled, the prior along with these parameters forms a posterior distribution over f.
We combined this with a hierarchical grid search  of the hyper-parameter space (especially for the composition of loops), and came up with values of the optimal parameters, given a particular input transform applied to datapoints. This is done only once per application. It should be emphasized here that the loops are never trained, only the RR algorithm which uses their outputs.

Note that the computational complexity of training the DLR is the complexity of the last stage (Ridge Regression), given the simple implementation of the loop.
\subsection{Ridge Regression}
The Ridge Regression (RR) model for the estimation of the weight coefficient matrix $W$ is calculated as
$$W=argmin_{W_{out}}\paren{\sum_{j=1}^{B}{\left\|y_j-W_{out}X\right\|_2^2+\lambda\left\|W_{out}X\right\|_2^2}}$$
where $B$ is number of training datapoints,  $y_j\in Y_{out}$ is the onehot device-label corresponding to the $j^{th}$ training datapoint, $W_{out}$ is the output weights to be trained, and $\lambda$ is the regularization factor. The RR equation can also be written in the closed form  $W = (X^T X +\lambda I_N )^{-1}  \ (X^T Y_{out} )$, where $X$ is the matrix of the B state vectors used for its training. 
\begin{figure}[h]
\vspace{-1mm}
\centering
\includegraphics[width=0.49\textwidth]{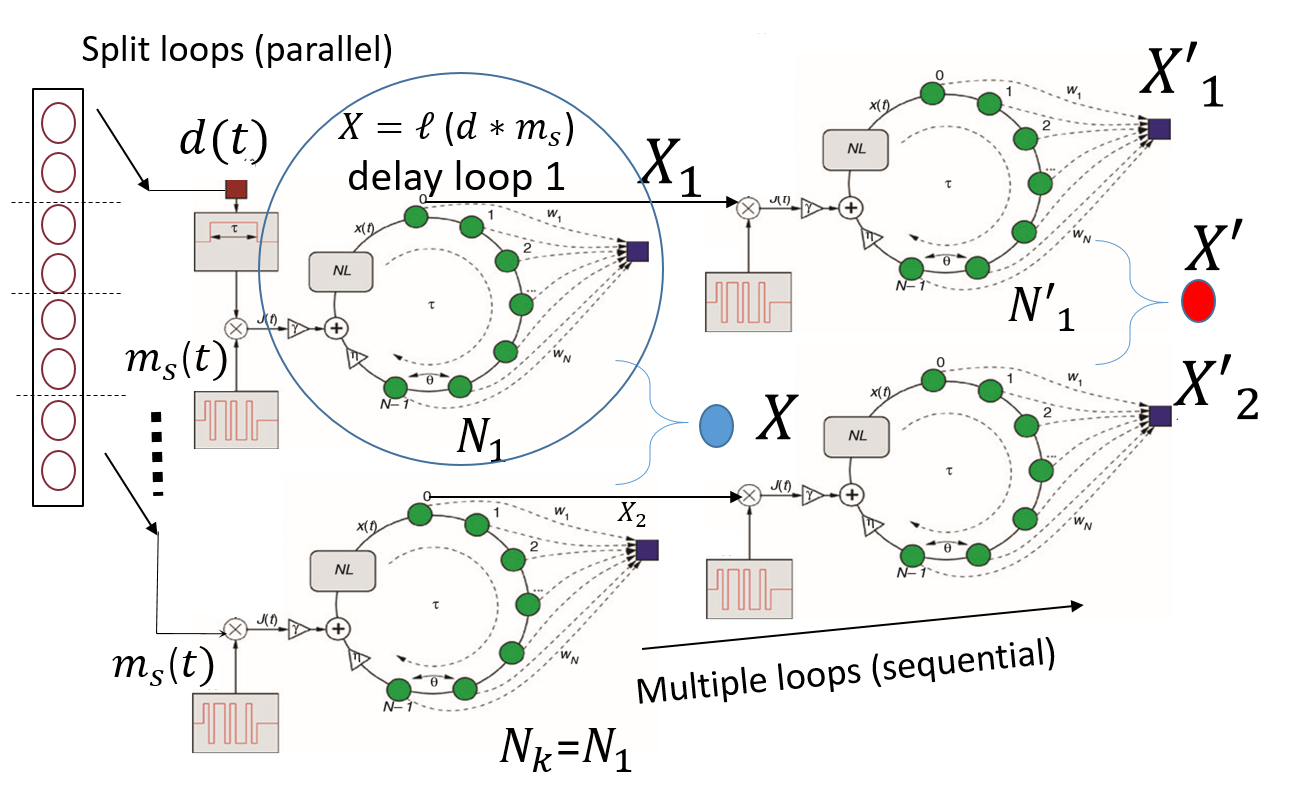}
\vspace{-6mm}
\caption{Using k split (parallel loops) reduces the size of the spreading mask and, hence, the size of each split loop while achieving the required projection into higher dimensional space. The joint state vector $X$ (marked by the blue circle) can be obtained as $\sum_{j=1}^{k}{X_j}$, or as a normalized scalar product of $X_j$s.}\vspace{-3mm}
\label{fig:multiple}
\end{figure}
\subsection{Split Loops}\label{subsec:split}
As Fig.~\ref{fig:multiple} shows, the split loops process the $k$ disjoint pieces of the split datapoint in parallel and result in the k state vectors $\curlb{X_1,\cdots, X_k}$. We can pass these $k$ outputs through another layer of parallel loops. The joint state vector $X$ (marked by the blue circle in Fig.~\ref{fig:multiple}, or red circle for a 2-layer design) can be obtained as $\sum_{j=1}^{k}{X_j}$, or as a normalized scalar product of $X_j$s. Both gave similar results in terms of the improved accuracy over a single loop. Let us emphasize here that DL is a dynamical system, and the reservoir size $N$ for a datapoint of size $\ell$ must be large enough to bring the DL into a dynamical state where the class separation happens.
Each of the split-datapoints is of size $\ell/k$ and its $N_j<N, j=1,\cdots,k$ now may be $k$ times smaller than the $N$ without splitting. Now, as the total $X$ is a linear combination and not concatenation of the split loops' $X_j$s, the entry to the RR is still of size $N_j$, which reduces the RR complexity from $BN^2$ to  $B(N/k)^2$. Beyond certain value of $k,$ the accuracy starts to drop as splitting the datapoint affects the samples that are no longer independent. The exact value of the threshold $k$ depends on the input transform applied to the burst of $I/Q$ samples composing the dataset. As an example, this threshold for FFT is above 10.  
The joint state vector results in higher accuracy than any of the $k$ split state vectors taken separately. For now it is important that the results obtained using the addition and scalar multiplication exceed the SoA, but we would also like to quantify the information loss if any. If we concatenated the split state vectors we would have preserved the information from the split datapoints completely but the resulting length of the state vector would have increased the complexity of learning by $k^2$. An information-theoretic analysis is under way on how much information is passed from the input to the state vector under different operations to construct the joint state.  
\subsection{Input Transforms}
It is known that certain transforms help losslessly compress the data if the information content is sparse, and some are more robust to noise. Motivated by this, we experimented with multiple input transforms in order to best mach the input to the architecture of DLR given constraints in the complexity (i.e., largest reservoir output we are willing to train). 
Since the RF waveforms are characterized by amplitude and phase we have 2 dimensions per sample in each 1024-long datapoint. Note that DLR loops cannot process complex-valued datapoints, and therefore all the transforms that we apply lose the phase information (except for the case in Fig.~\ref{fig:mix}) 
\subsubsection{Complex Amplitudes}
\begin{figure}[h]
\vspace{-1mm}
\centering
\includegraphics[width=0.39\textwidth]{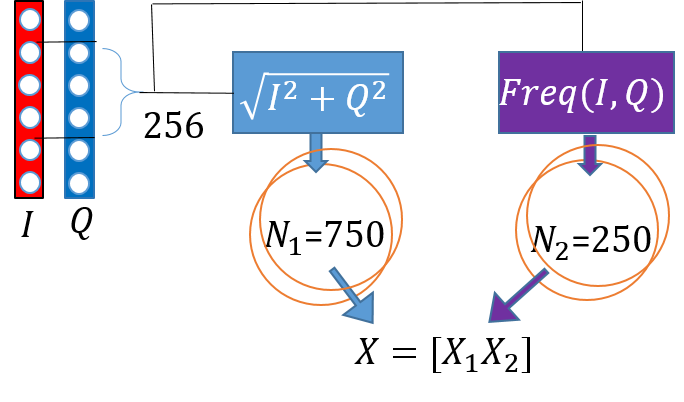}
\vspace{-5mm}
\caption{Using diverse transforms showed some initial promise in combining complementary transforms - magnitudes are combined with simple frequency estimates}\vspace{-3mm}
\label{fig:mix}
\end{figure}

We consider the transform from complex-valued samples to their amplitudes as the baseline. Here, subbursts of 256 complex amplitudes are extracted to replace the longer datapoints of 1024-samples. For this extraction we used a simple linear classifier to explore how salient information is distributed along the datapoint. We conducted a systematic experiment in which we have been replacing each datapoint with a subset of contiguous complex-valued samples within that datapoint, and evaluating the accuracy of the classifier for different sizes of such subsets, and different offsets from the start. The rationale was that if the salient information was concentrated in a smaller sub-burst, than the size of the reservoir $N$ can also be scaled down while maintaining the same accuracy.  The 256 samples per datapoint, extracted at a location that preserves the accuracy achieved with the original burst, are equivalent to a 1us long burst of samples. 
The accuracy of DLR based on the amplitudes of the subburst of 256 complex samples ranks the worst among the tested transforms. Please observe how the accuracy for 2 split loops in Fig.~\ref{fig:acc} gets higher with the reservoir size when amplitudes (blue - plain vanilla transform) are replaced by the amplitudes of complex FFT transforms. We have also combined the amplitudes in one pair of split loops with the frequency estimates in another and concatenated the state vectors. With the loops processing the magnitudes of dimension 750, and the loops processing the frequency estimates of dimension 250 (as in Fig.~\ref{fig:mix}),  we get the total state vector of length N=1000 and the accuracy of 93.8\%. Note that frequency estimates \cite{SKay} of the complex valued input bursts are real valued bursts of smaller length than the original bursts, as we are using 3 consecutive I/Q samples to calculate one frequency value. This approach should be further explored with a finer frequency estimate, and exhaustive loop combinations.
\subsubsection{FFT}
The exploration of information salience above revealed the sparsity of the original dataset that can be leveraged through additional transforms. We next optimized the performance of the digital test bench when the input to the loop are magnitudes of the FFT transform applied to the bursts of complex valued RF samples. The initial findings were that the FOMs deteriorate since the FFT transform ‘spreads’ the information across the datapoint. Specifically, without FFT, the salient information is limited to the subburst of length 256 located in the middle of the 1024 long burst of sampled signal magnitudes. When the FFT magnitudes are used as datapoints, the salient information is located in all 1024 samples. This requires a delay loop longer than 1024, exceeding the loop size of 600 or 800 which produced the best accuracy-complexity trade-offs.
\begin{figure}[h]
\centering
\hspace{-4mm}\includegraphics[width=0.52\textwidth]{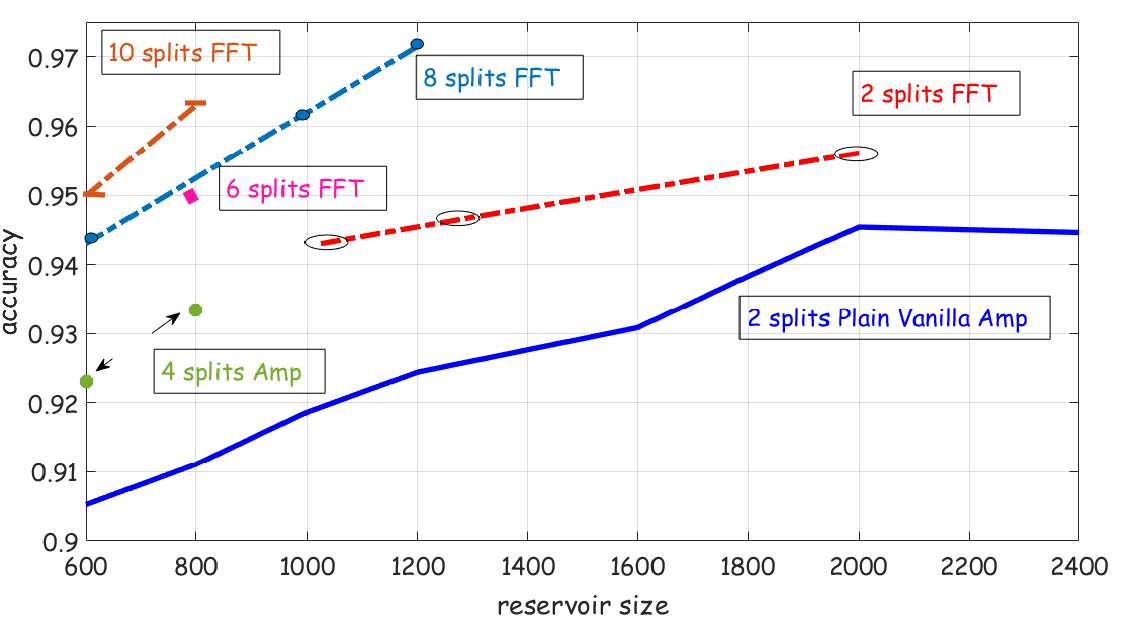}
\vspace{-3mm}
\caption{Accuracy as a function of the reservoir size for split loops when all using the same input transforms - either signal amplitudes and FFT amplitudes. Notice the improvements with 10 splits, unchanged complexity.
}\vspace{-4mm}
\label{fig:acc}
\end{figure}
However, using the technique of loop splitting, we were able to utilize 8-fold splits to decrease the input dimension, and consequently reduced the reservoir size in each of the loops  to the previously used values. While this kept the H/W reduction factors unchanged, we achieved a gain in accuracy. Please see Fig.~\ref{fig:acc}. Note that splitting the datapoints based on the complex signal amplitudes into 8 loops deteriorates the accuracy compared to 4 loops, which makes the FFT more effective in achieving the best accuracy-complexity trade-off.
\subsubsection{Differential FFT}
We here find the average of the burst amplitudes across the training dataset and remove it from each datapoint amplitude, then create new complex-valued bursts with  the common bias so removed (while preserving the phase). We perform the FFT of the resulting waveform, referring to its magnitudes as the {\em differential FFT} transform. This transform helped increase the robustness of the photonic loop by reducing its sensitivity to the in-loop noise: compared to the regular FFT magnitudes, differential FFT increased the accuracy by more than 15\% for the same reservoir size. In the digital loop, differential FFT did not make a difference.
\subsubsection{Decimated DFT}   
\begin{figure}[h]
\centering
\hspace{-2mm}\includegraphics[width=0.5\textwidth]{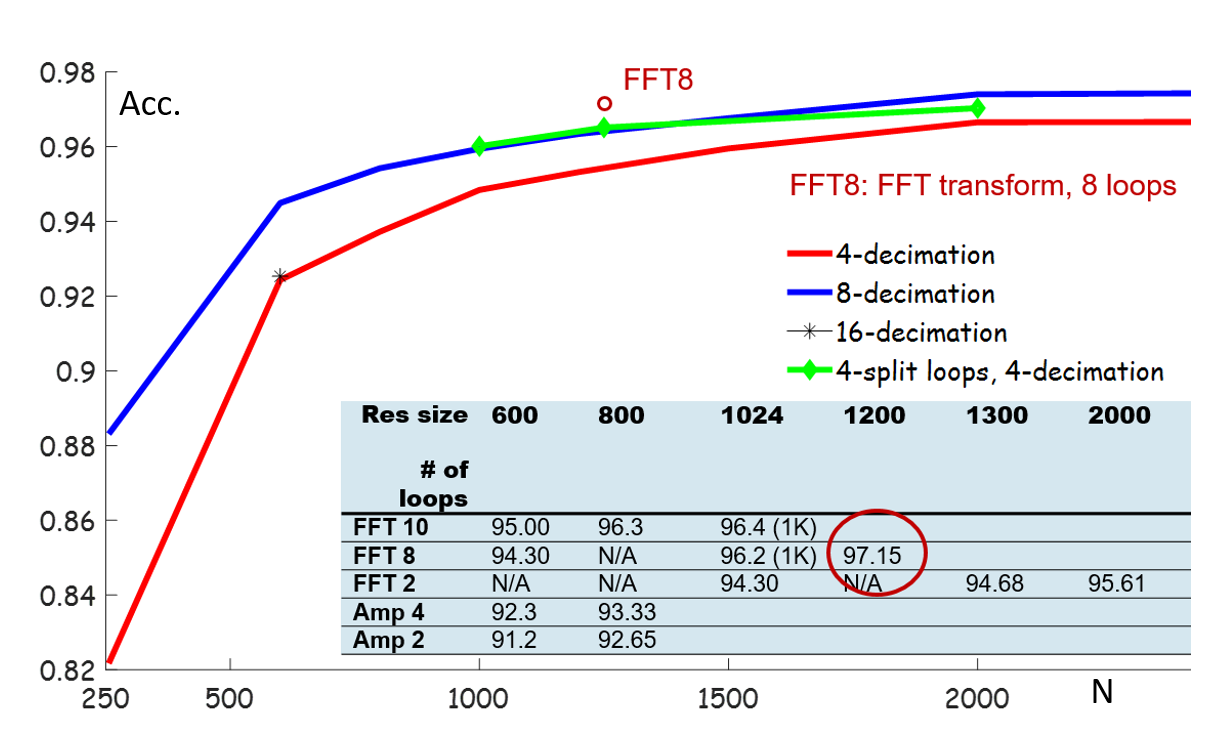}
\vspace{-3mm}
\caption{Accuracy vs. N with decimation of DFT: N= 1200 achieves 97.48\% of accuracy with 2 split loops and $d=8,$ or 4 split loops and $d=4.$
With further decimation (16-fold) the accuracy falls back to 4-fold decimation (2 loops).
}\vspace{-5mm}
\label{fig:accDFT}
\end{figure}

We decimate the DFT matrix 
\begin{equation}   \label{eq:DFT}
    D = \frac{1}{N}
    \left( \begin{array}{ccccc}
        1 &  1 & 1 & \ldots & 1 \\
        1 & \omega_N^{-1} & \omega_N^{-2} & \ldots & \omega_N^{-(N-1)} \\
        1 & \omega_N^{-2} & \omega_N^{-4} & \ldots & \omega_N^{-2(N-1)} \\
        \vdots&\vdots&\vdots& &\vdots\\
        1 & \omega_N^{-(N-1)} & \omega_N^{-2(N-1)} &\ldots&\omega_N^{-(N-1)^2} 
    \end{array} \right),
\end{equation}
where $\omega_N = e^{i 2\pi/N} \in \C$, by some factor $d$ - by keeping every $d^{th}$ column of $D$ - creating a sparse DFT matrix $D_d$ of size 1024 by 1024/d. By multiplying the complex-valued burst of size 1024 by $D_d$ we obtain a decimated discrete frequency transform whose length is reduced $d$ times. We use its amplitudes as datapoint to split into 2 parallel loops. Please see Fig.~\ref{fig:accDFT} where the accuracy is shown for different  reservoir sizes.  If we use more split loops the decimation factor must be reduced to maintain the accuracy. Given a compressive transform, we observe that the compression affects the number of split loops needed to achieve accuracy gain for the same reservoir size. Split loops become useless after a certain point, which depends on the input transform.
Given this trade-off, an interesting future goal is to compare the cost of performing per-datapoint transform vs splitting the loop.
\section{Discussion of Experimental Results}
\begin{figure}[h]
\vspace{-1mm}
\centering
\hspace{-1mm}\includegraphics[width=0.5\textwidth]{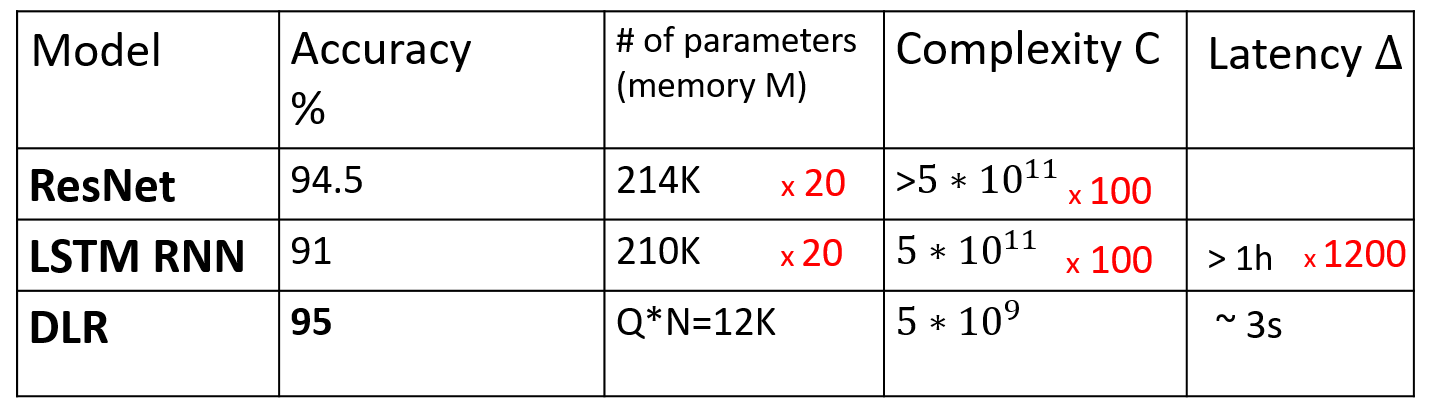}
\vspace{-3mm}
\caption{Figures of Merit compare the best trade-off of {em SEI accuracy vs H/W reduction} for DLR in comparison with SoA. Complexity is expressed in terms of total MAC operations and memory in terms of the number of parameters to train.
}\vspace{-2mm}
\label{fig:fom}
\end{figure}

The table of FOMs in Fig.~\ref{fig:fom} shows the 20-device SEI performance numbers for the best trade-off we achieved with DLR. These are compared with the equivalents for SoA neural networks trained on a single GPU. The red numbers show the reduction factor obtained with DLR compared to respective values for ResNet or RNN. The spatial reduction factor for DLR vs SoA is the ratio of the number of trainable parameters (20), and the power reduction factor is the ratio of the complexity of training (100), while the latency compares the hours of training versus 3 seconds achieved with our demo platform ($\geq 1200$). For the calculation patterns of FOMs, please see our paper on intermediate results \cite{GomacTech}.The result here use the reservoir size of 600. Note that we could achieve higher accuracy  of above 97\% by using the reservoir size of 1200, which increases the complexity 4 x (consult Fig.~\ref{fig:acc}).

Finally, we would like to present the results obtained using the same loop hyper-parameters (without retuning) for the application of wireless protocol recognition. We here achieved 99\% accuracy in classification of the 4 ISM protocols trained on clean signal collections. We collected data from 5 devices per protocol, 4 from a common manufacturer, the 5th from a different one. We either used connectorized devices in conjunction with cables,
circulators and attenuators or an RF shield box to ensure pristine (clean) RF data. We used the same USRP and the same sample rate as for the SEI dataset. After the signals were basebanded and the bandwidth information removed, this accuracy dropped by about 2 \%. Removing BW information emulates the efforts to hide the identity by transmitting on a different frequency and performing rate adaptation. 

\begin{figure}[h]
\centering
\hspace{-4mm}\includegraphics[width=0.54\textwidth]{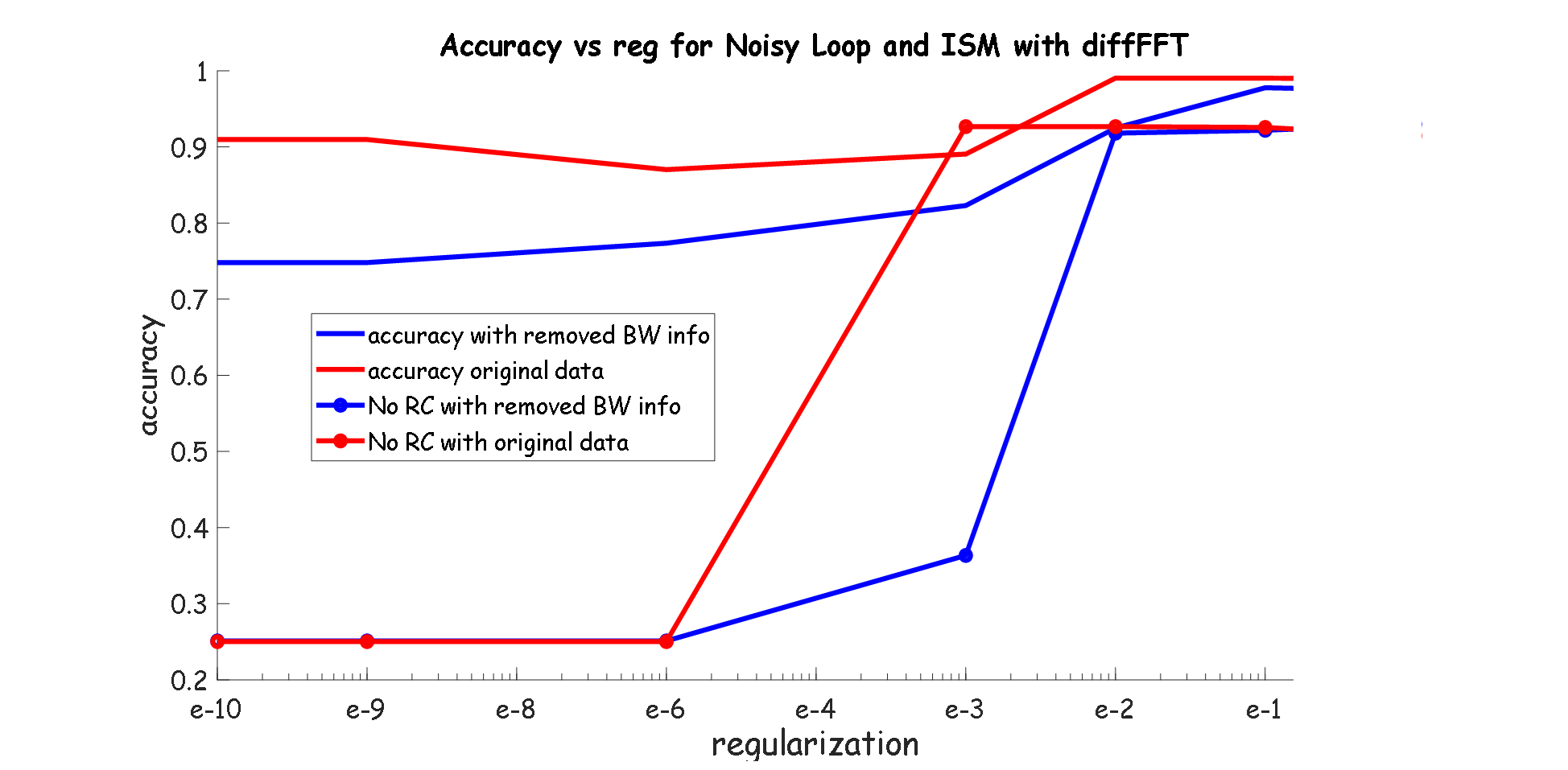}
\vspace{-3mm}
\caption{Accuracy vs RR regularization for classification of the 4 ISM protocols with and without BW as salient feature.
}
\label{fig:ism}
\end{figure}

Fig.~\ref{fig:ism} shows both cases on DLR (plots without markers), as well on the simple ridge regression (without reservoirs) for the same complexity. The simple RR has a drop in accuracy of about 9\%.  An interesting observation is that the results stayed the same after the loop noise has been added, modeled from the photonic loop observations. We confirmed this result from the digitally implemented loop by performing the classification in our photonic loop. DLR also produces a better conditioned weight matrix, i.e. the results are more stable for different values of the regularization parameter $\lambda$ than the RR without the reservoir (see Fig.~\ref{fig:ism}). The results  with other transforms are consistent, and we omit them for the lack of space. The idea here was to test if DLR can be re-purposed for other signals, and the results exceeded our expectations as we did not have to re-optimize the parameters. 

Admittedly, both the ISM dataset and the 4-way classification based on it are much less complex. However, comparing it with our published worked \cite{SDAE} that trained an ISM classifier off-line in order to minimize the complexity of inference, we ended up with 4 times smaller inference complexity, and evidently much simpler and faster training. Future work should include signal corruptions by fading channels and interference.
\section{Conclusion}
We demonstrated the classification accuracy, real-time efficiency and energy reduction of our novel design for in-situ training and inference in two RF spectrum monitoring applications using innovative {\em delay-loop reservoir design and architectures}. Our delay-loop (DL) design, as an efficiently implemented random
recurrent topology, prepares the training data sequences to successfully train linear classification algorithms. In particular, using real devices we showed that Specific Emitter Identification (SEI) for secure IoT authentication can be trained orders of magnitude faster on compact platforms achieving  remarkable accuracy. This is very important for vulnerable and resource-constrained IoT devices. We presented algorithms for both photonic and digital loops in a closed form, and gave an intuitive explanation how a single neuron in  the delay-loop replaces the N neurons in a traditional spatial reservoir (where these neurons must be connected via recurrent connections). While the presented prototype platform uses both digital and photonic realizations of our design of delay loops, our performance figures are based on the former and reflect architectures with trees of DLs. Combining the splitting and multi-layer loops, with different degrees of asymmetry between inner and outer DLs, we achieved various levels of  trade-off between the accuracy and complexity of DLR.  The state of all loops in the tree are read out at each input datapoint, and then concatenated or combined to serve as classification datapoints. We described the effects of architectural loop combinations and their interplay with various input transforms. By applying information-theoretic measures on a quantized version of DLR, we continue to further study these effects as an end-to-end information transfer. Future work includes new applications and additional exploration of loop architectures including when every DL is supporting a different data transform.

{\bf ACKNOWLEDGMENT: }This research was funded by DARPA. The views and conclusions contained in this document are those of the authors and should not be interpreted as representing the official policies, either expressed or implied, of the U.S. Government. 

DISTRIBUTION STATEMENT A. Approved for public release: distribution unlimited
\bibliographystyle{IEEEtran}%
\bibliography{DLRbasic}
\end{document}